\documentclass[twocolumn,aps,pra]{revtex4-1}
\usepackage{epsfig}
\usepackage[english]{babel}
\usepackage{latexsym}
\usepackage{subfigure}
\usepackage{graphics}
\usepackage{epstopdf}
\usepackage{dcolumn}
\usepackage{amsmath}
\usepackage{hyperref}
\usepackage{amssymb}
\usepackage{appendix}
\usepackage{color}
\usepackage{lineno}
\usepackage{soul}
\usepackage{booktabs}
\usepackage{longtable}
\usepackage{multirow}
\usepackage{array}
\usepackage{ulem}

\begin{document}

\title{Effect of Berry connection on attosecond transient absorption spectroscopy in gapped graphene}
\author{Jiayu Yan$^{1}$, Hongxuan Er$^{1}$, Wei Dang$^{1}$, Jiahuan Ren$^{1}$, Chao Chen$^{2,\dag}$, Dianxiang Ren$^{3}$, Fulong Dong$^{1,*}$}

\date{\today}

\begin{abstract}

We investigate the attosecond transient absorption spectroscopy (ATAS) in gapped graphene by numerically solving the four-band density matrix equations.
Our results reveal that, in contrast to pristine graphene, whose fishbone-shaped spectra oscillate at twice the pump laser frequency, the ATAS of gapped graphene exhibits an additional component oscillating at the pump laser frequency, induced by the Berry connection.
To gain insight into these interesting results, we employ a simplified model to derive an analytical expression for the spectral component stemming from the Berry connection.
Our analytical results qualitatively reproduce the key features observed in the numerical simulations, revealing that the intensity of the fundamental-frequency spectral component depends not only on the Berry connection but also on the energy shifts associated with the effective mass of electrons at the van Hove singularities.
These results shed light on the complex generation mechanism of the ATAS in symmetry-broken materials.

\end{abstract}
\affiliation{
$^{1}$College of Physics Science and Technology, Hebei University, Baoding 071002, China\\
$^{2}$College of Physics and Electronic Engineering, Xingtai University, Xingtai 054001, China\\
$^{3}$Shanxi Institute of Energy, Taiyuan 030600, China}

\maketitle

\section{Introduction}

Attosecond transient absorption spectroscopy (ATAS) provides a powerful tool to probe ultrafast electronic dynamics on the attosecond timescale \cite{Eleftherios,MetteBGaarde,AnneliseRBeck,MengxiWu}.
Combining an attosecond pulse with a strong pump laser, it provides an all-optical method to explore light-matter interactions, possessing high temporal and spectral resolution.
The ATAS has been successfully applied to study electron dynamics of atoms and molecules \cite{HeWang,MHoller,ZQYang,ShaohaoChen,MichaelChini,PPeng1,PPeng2,DTMatse,AKaldun,SBaker,SSeverino}.

In recent years, significant progress has been made in applying ATAS to bulk solids \cite{MartinSchultze,MLucchini,FSchlaepfer,MVolkov,TOtobe,MatteoLucchini,HMashiko,HMashikoK} and various two-dimensional materials \cite{KUchida,GioCistaro, Dong4}.
An interesting fishbone-shaped structure has been observed in the ATAS for the periodic materials \cite{MLucchini,TOtobe1,SYamada,Dong4}.
One of the typical characteristics of this structure is that its oscillation frequency is observed to be twice the frequency of the pump laser \cite{Dong4,MVolkovSato,SYamadaYabana,YKim}.
More recently, an analytical study on the ATAS of graphene \cite{Dong4} has revealed that
the spectra are dominated by intraband electron dynamics, i.e., ``dynamical Franz-Keldysh effect" \cite{MLucchini,TOtobe1}.
More specifically, the generation of the fishbone structure is closely related to the band structure, particularly to the effective mass of electrons at the van Hove singularities.

On the other hand, an important concept capturing the topological nature of Bloch electrons is the Berry connection, a gauge-dependent vector potential that plays a crucial role in dominating electron dynamics under external fields \cite{DXiao}.
Although its curl, the Berry curvature, has attracted significant attention for its role in phenomena such as the anomalous Hall effect \cite{NNagaosa} and topological charge pumping \cite{XLQi,ZSun}, the direct physical consequences of the Berry connection itself remain less well understood, especially in the context of strong-field and ultrafast dynamics.
In contrast to symmetry materials like graphene, the Berry connection may play a significant role in intraband electron dynamics in symmetry-broken crystals, thereby shaping the features of the ATAS.

As a symmetry-broken two-dimensional material, gapped graphene has recently attracted significant attention \cite{SAOMotlagh,MYu,FPakdel}.
Owing to possessing its nontrivial Berry connection, it provides a promising platform for exploring the influence of the Berry connection on the ultrafast absorption spectroscopy.
In this work, we numerically solve the density matrix equations based on a developed two-dimensional four-band model of gapped graphene to simulate the ATAS.
In contrast to pristine graphene, whose fishbone-shaped spectra primarily oscillate at twice the pump laser frequency, the numerical spectra exhibit a spectral component oscillating at the frequency of the pump laser, which can be attributed to the presence of Berry connections in gapped graphene.
To gain further insight into this interesting spectral component, we approximate the two-dimensional four-band structure using a simplified model to derive an analytical expression for the spectral component induced by the Berry connection.
Our analytical results indicate that the intensity of the fundamental-frequency spectral component (FFSC) depends not only on the Berry connections but also on the energy shifts associated with the effective mass of electrons at the van Hove singularities.

This paper is organized as follows.
We describe our numerical simulation methods and results in Sec. \ref{s2}.
We give the Analytical study for the effect of the Berry connection on the ATAS in Section \ref{s3}.
Finally, Sec. \ref{s4} presents our conclusion.

\section{Numerical simulation methods and results}

\label{s2}

\subsection{Four-band structure of gapped graphene}

As shown in Fig. 1(a), graphene has a honeycomb lattice composed of two sublattices, labeled ``A" and ``B" \cite{Castro}.
The lattice vectors $\boldsymbol{a}_{1}$ and $\boldsymbol{a}_{2}$, as well as the carbon-carbon bond length $d= 1.42$ \AA $ $ are also indicated.
Figure 1(b) illustrates the first Brillouin zone of gapped graphene.
In this work, we consider four energy bands : two core bands, denoted as $g_{1}$ and $g_{2}$, as well as the valence ($v$) and conduction ($c$) bands \cite{Dong4}.
Utilizing the Bloch states as the basis set, the tight-binding Hamiltonian $H_{0}$ takes the form
$ H_{0}=\left(\begin{array}{cc}
\Delta_g/2 & \gamma_{0} f(\mathbf{k}) \\
\gamma_{0} f^{*}(\mathbf{k}) & -\Delta_g/2
\end{array}\right)$,
where electrons are allowed to hop only between nearest-neighbor atoms, with a hopping energy $\gamma_{0}=0.1$ a.u. (Throughout the paper, atomic units are used if not specified.)
The structure factor is given by $f(\textbf{k})=e^{i \texttt{k}_{x}d}+2\cos( \sqrt{3}\texttt{k}_{y}d / 2)e^{-i\texttt{k}_{x}d/2}$.
Diagonalizing $H_{0}$ yields energy eigenvalues for the $c$ and $v$ bands as $\varepsilon_{c}(\textbf{k}) = -\varepsilon_{v}(\textbf{k}) =\sqrt{\gamma_{0}^2 \vert f(\textbf{k}) \vert^2 + \Delta_g^2/4}$ which describe the electronic dispersion near the band gap.
The dispersion relations for the two core bands are given by $\varepsilon_{g_1} = \varepsilon_g - \Delta_g/2$ and $\varepsilon_{g_2} = \varepsilon_g + \Delta_g/2$ with $\varepsilon_g = -280$ eV.
These bands are depicted in Fig. 1(c) (see Sec. I of the Supplemental Material for further details).

\subsection{Four-band density matrix equations}

We numerically simulate the ATAS of gapped graphene using the four-band density matrix equations \cite{Dong4}.
Within the dipole approximation, these equations are

\vspace{-0.4cm}
\begin{align}
 &i \dfrac{d}{d t}\rho_{mn}(\textbf{\textrm{k}}_t,t,t_{d}) = [ \varepsilon_{mn}(\textbf{\textrm{k}}_t) - i \Gamma_{mn}]\rho_{mn}(\textbf{\textrm{k}}_t,t,t_{d}) + \nonumber\\
& \textit{\textbf{E}}(t, t_{d}) \cdot
\sum_{l} [\textbf{\textrm{D}}_{ml}^{\textbf{\textrm{k}}_t}  \rho_{ln}(\textbf{\textrm{k}}_t,t,t_{d}) - \rho_{ml}(\textbf{\textrm{k}}_t,t,t_{d}) \textbf{\textrm{D}}_{ln}^{\textbf{\textrm{k}}_t}],
\end{align}
where $\textbf{k}_t = \textbf{k} + {\textit{\textbf{A}}}_{I}(t, t_d)$ is the crystal momentum in the presence of the vector potential, with $\textbf{k}$ taken from the first Brillouin zone.
The energy difference is defined as $\varepsilon_{mn}(\textbf{k})=\varepsilon_{m}(\textbf{k})-\varepsilon_{n}(\textbf{k})$, in which $m, n \in \{g_{1}, g_{2}, v, c\}$.
We set the relaxation parameters as follows: $\Gamma_{mn}$ are $\Gamma_{g_1v}=\Gamma_{g_1c}=\Gamma_{g_2v}=\Gamma_{g_2c}=0.004$ a.u. $\equiv \Gamma_{0}$  \cite{GioCistaro}, while the remaining terms satisfy $\Gamma_{g_1g_1}=\Gamma_{g_2g_2}=\Gamma_{cc}=\Gamma_{cv}=\Gamma_{vv}=\Gamma_{g_1g_2}=0$.
The transition dipole matrix elements $\textbf{\textrm{D}}_{mn}^{\textbf{\textrm{k}}}$ are discussed in detail in Sec. I of the Supplemental Material.

\begin{figure}[t]
\begin{center}
{\includegraphics[width=8.5cm,height=7cm]{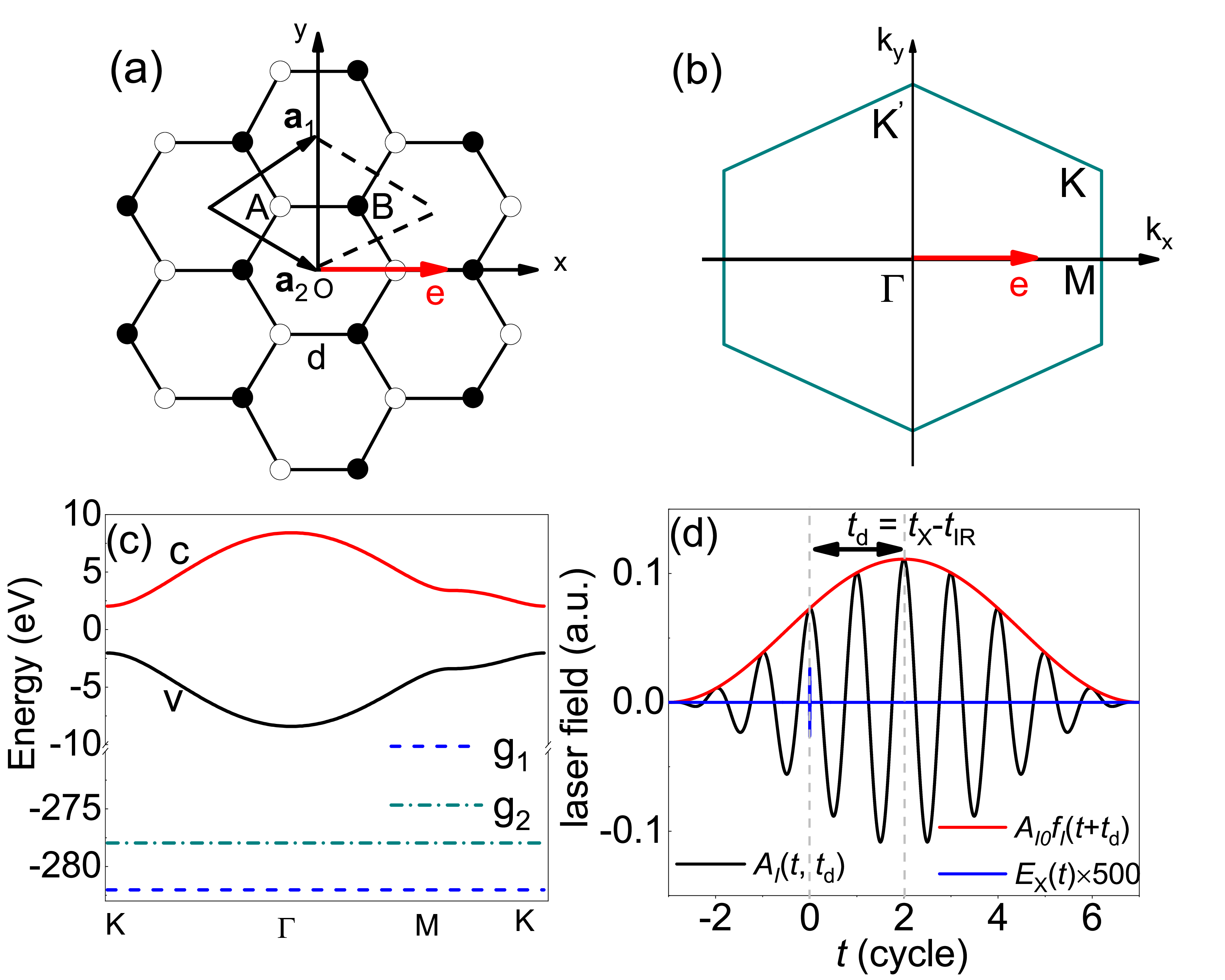}}
\caption{
(a) Hexagonal lattice structure of two-dimensional gapped graphene.
Each primitive cell contains two atoms labeled ``A" and ``B".
$\boldsymbol{a}_{1}$ and $\boldsymbol{a}_{2}$ are the lattice vectors.
$\boldsymbol{e}$ is the unit vector that indicates the polarization direction of the electric field.
$d$ is the carbon-carbon distance.
(b) First Brillouin zone of gapped graphene with high symmetry points $\Gamma$, $\textsc{M}$, and $\textsc{K}$.
(c) Two-dimensional four-band structure of gapped graphene.
(d) Schematic of the time delay between the IR pump laser and the X-ray probe pulse.
}
\label{fig:graph1}
\end{center}
\end{figure}

In Eq. (1), the total electric field is given by $\textit{\textbf{E}}(t, t_{d}) = \textit{\textbf{E}}_{I}(t, t_{d}) + \textit{\textbf{E}}_{X}(t) $,
where $\textit{\textbf{E}}_{X}(t)$ represents the electric field of the X-ray pulse, defined as
$\textit{\textbf{E}}_{X}(t) =\textit{E}_{X} f_{X}(t) \cos(\omega_{X}t) \textit{\textbf{e}}_{z}$ with a Gaussian envelope $f_{X}(t) = e^{-(4 ln 2)(t/ \tau_{X})^{2}}$, corresponding to a full width at half maximum of $\tau_{X} = 80$ attoseconds.
The peak amplitude $\textit{E}_{X}$ corresponds to an intensity of $1 \times 10^{8}$ W/cm$^{2}$, and the X-ray frequency is set to $\omega_{X} = 280$ eV.
Here, $\textit{\textbf{e}}_{z}$ is the unit vector perpendicular to the graphene monolayer.

The electric field of the infrared (IR) laser is calculated by $\textit{\textbf{E}}_{I}(t, t_d) = -{\partial \textit{\textbf{A}}_{I} (t, t_d)}/{\partial t} $, in which $\textit{\textbf{A}}_{I}(t, t_d) = A_{I0} f_{I}(t+t_d) \cos(\omega_{I} t+\omega_{I} t_d) \textit{\textbf{e}}$ is the vector potential of the IR laser field, as shown in Fig. 1(d).
The envelope function is defined as $f_{I}(t) = \cos^2 (\omega_{I}t / 2n)$ with $n = 10$.
The amplitude $\textit{A}_{I0}$ corresponds to a laser intensity of $1 \times 10^{10}$ W/cm$^{2}$.
The laser frequency $\omega_I$ corresponds to a wavelength of $\lambda = 3000$ nm.
$T=2 \pi / \omega_I$ is the period of the IR laser field.
The time delay between the two pulses is given by $t_d = t_{X} - t_{IR}$, where $t_{X} = 0$ and $t_{IR}$ represent the peak times of the X-ray and IR laser envelopes, respectively.
When $t_d = 0$, the peaks of both pulses coincide in time.
$\textit{\textbf{e}}$ is the unit vector along the $\Gamma - \textsc{M}$ direction of gapped graphene, as shown in Figs. 1(a) and 1(b).

Equation (1) is numerically solved using the standard fourth-order Runge-Kutta algorithm.
The X-ray response function is calculated by \cite{MetteBGaarde}

\vspace{-0.4cm}
\begin{subequations}
\begin{align}
S^{X}(\omega) = &2 \operatorname{Im}[\int^{t}_{0}r^{X}(t)e^{-i\omega t}dt],\\
S(\omega,t_d) = &2 \operatorname{Im}[\int^{t}_{0}r(t,t_d)e^{-i\omega t}dt],\\
S^{\boldsymbol{\mathcal{A}}=0}(\omega,t_d) = &2 \operatorname{Im}[\int^{t}_{0}r^{\boldsymbol{\mathcal{A}}=0}(t,t_d)e^{-i\omega t}dt].
\end{align}
\end{subequations}
Here, the total dipole is given by $r(t,t_d) = \sum_{\textbf{k}} \sum_{i,g} r_{gi}(\textbf{k},t,t_d)
= \sum_{\textbf{k}} \sum_{i,g}[\textrm{D}^{\textbf{k}_t}_{gi} \rho_{ig}(\textbf{k}_t,t,t_d) + c.c.]$, where $g \in \{g_1, g_2\}$ and $i \in \{v, c\}$.
The term $S^{X}(\omega)$ represents the X-ray-only response function, obtained by evaluating the dipole $r^{X}(t)$ in the absence of the IR field.
The function $S(\omega, t_d)$ denotes the full function at time delay $t_d$.
To isolate the contribution of the Berry connections to the ATAS, we calculate $S^{\boldsymbol{\mathcal{A}}=0}(\omega,t_d)$ by artificially setting that $\textbf{D}^{\textbf{k}}_{cc}$, $\textbf{D}^{\textbf{k}}_{vv}$, $\textbf{D}^{\textbf{k}}_{g_1g_1}$, and $\textbf{D}^{\textbf{k}}_{g_2g_2}$ are zero in the process of solving Eq. (1).
Note that although $\textbf{D}^{\textbf{k}}_{nn}$ with $n \in \{ c, v, g_1, g_2 \}$ is gauge dependence, the difference $\boldsymbol{\mathcal{A}}_{cg}^{\textbf{k}} = \textbf{D}^{\textbf{k}}_{cc} - \textbf{D}^{\textbf{k}}_{gg}$ with $g \in \{g_1, g_2 \}$ is gauge invariant in our model.

The ATAS is then evaluated using

\vspace{-0.4cm}
\begin{subequations}
\begin{align}
\Delta S(\omega,t_{d}) = &S(\omega,t_{d}) - S^{X}(\omega),\\
\Delta S^{\boldsymbol{\mathcal{A}}}(\omega,t_{d})=& S(\omega,t_{d})- S^{\boldsymbol{\mathcal{A}}=0}(\omega,t_{d}).
\end{align}
\end{subequations}
Here, $\Delta S^{\boldsymbol{\mathcal{A}}}(\omega,t_{d})$ captures the spectral modifications induced by the Berry connections.
To analyze the oscillation frequency of the spectra, we calculate the energy-frequency maps using
\vspace{-0.4cm}
\begin{subequations}
\begin{align}
\Delta \tilde{S}(\omega,N) =& \left\vert \int \Delta S(\omega,t_{d}) e^{-iN \omega_0 t_d} d t_d \right\vert^2  ,\\
\Delta \tilde{S}^{\boldsymbol{\mathcal{A}}}(\omega,N) =& \left\vert \int \Delta S^{\boldsymbol{\mathcal{A}}}(\omega,t_{d}) e^{-iN \omega_0 t_d} d t_d \right\vert^2.
\end{align}
\end{subequations}
Here, $N$ denotes the spectral oscillation frequency.

\subsection{Numerical results for the ATAS}

\begin{figure}[t]
\begin{center}
{\includegraphics[width=8.5cm,height=14cm]{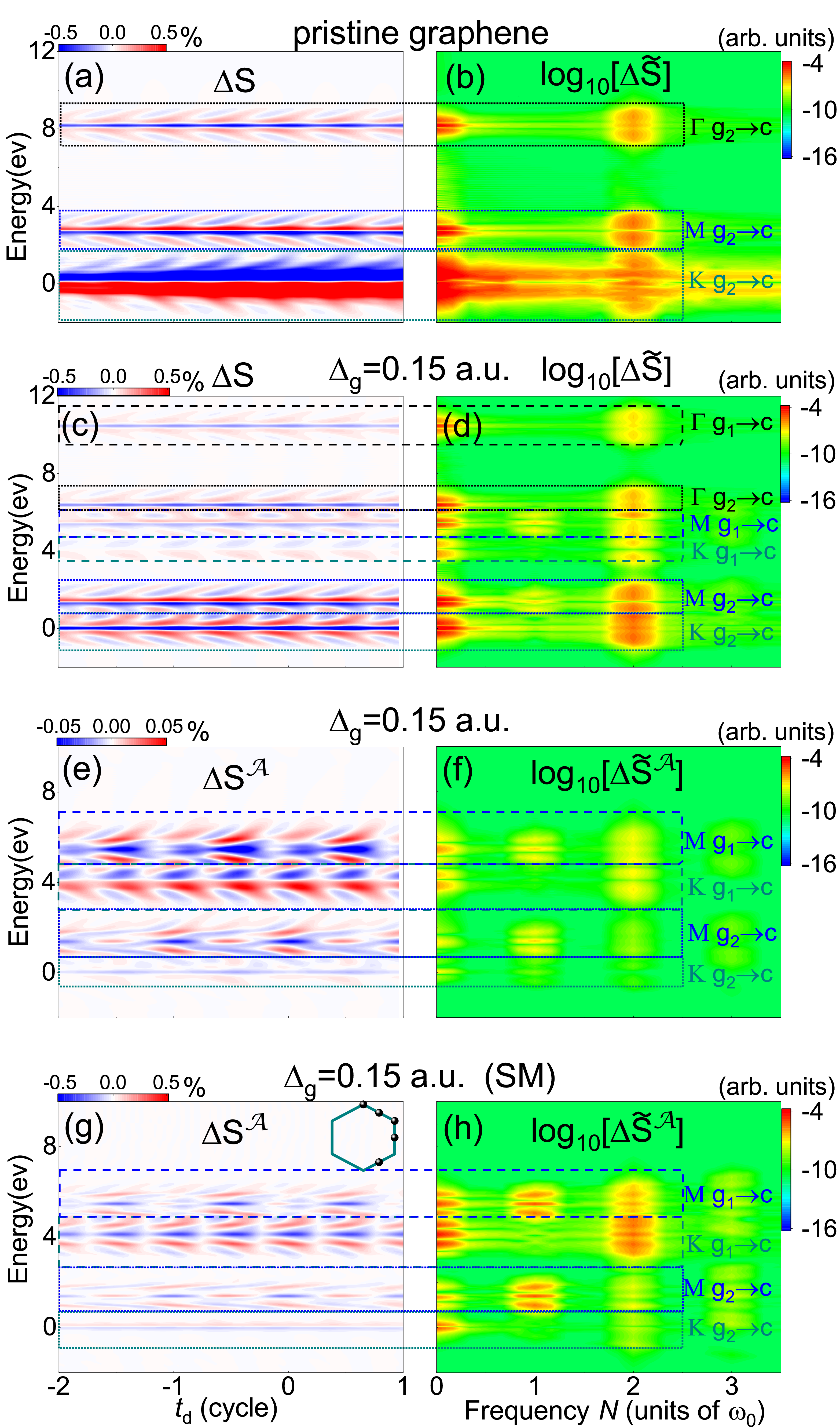}}
\caption{
(a) ATAS of pristine graphene as a function of the time delay in units of IR laser cycles, which is calculated using Eq. (3a).
(b) Frequency-energy map calculated by Eq. (4a) corresponding to the ATAS in (a).
(c), (d) Same as (a) and (b), respectively, but for gapped graphene with the energy gap of $\Delta_g = 0.15$ a.u.
(e) Spectra $\Delta S^{\boldsymbol{\mathcal{A}}}(\omega,t_{d})$ calculated by Eq. (3b).
(f) Frequency-energy map calculated by Eq. (4b), corresponding to the spectra in (e).
(g), (h) Same as (e) and (f), respectively, but the results are numerically calculated based on the simplified model (SM) as shown in the inset.
In panels (a)-(h), the rectangles label the spectra related to the transition $g_1 \rightarrow c$ (or $g_2 \rightarrow c$) for electrons located at $\Gamma$, $\textrm{M}$, and $\textrm{K}$ points, respectively.
}
\label{fig:graph1}
\end{center}
\end{figure}

By numerically solving the four-band density matrix equations, we calculate the ATAS $\Delta S(\omega,t_{d})$, as defined in Eq. (3a), for pristine graphene.
The resulting spectra are shown in Fig. 2(a).
Here, the spectra are plotted as a function of the time delay $t_d$, expressed in units of the IR laser optical cycles.
Noted that the ATAS has been normalized by $S^{X}(\omega)$, and the same normalization is applied in subsequent figures.
In pristine graphene, optical transitions between the $g_1$ and $c$ bands are forbidden. As a result, the spectra observed in Fig. 2(a) originate from the electron transitions from the $g_2$ band to the $c$ band ($g_2 \rightarrow c$), induced by X-ray pulse \cite{Dong4}.
Around the spectral energies $\varepsilon_c(\textbf{k}_{\textrm{M}})$ and $\varepsilon_c(\textbf{k}_{\Gamma})$, the spectra exhibits fishbone structures, whose generation mechanisms have been discussed in details in Ref. \cite{Dong4}.
Near the spectral energy $\varepsilon_c(\textbf{k}_{\textrm{K}})$, the electrons excited by the IR laser from the $v$ to $c$ bands suppress the electron transitions from the $g_2$ to $c$ bands induced by the X-ray pulse.
Simultaneously, resulting holes in the $v$ band can receive electrons arriving from the $g_1$ band, forming the spectrum structure near the $\textrm{K}$ point in Fig. 2(a) \cite{Dong4,GioCistaro}.

Corresponding to the ATAS shown in Fig. 2(a), the energy-frequency map $\textrm{log}_{10}[\Delta \tilde{S}(\omega,N)]$ calculate by Eq. (4a) is displayed in Fig. 2(b).
The map reveals that the fishbone structures around $\varepsilon_c(\textbf{k}_{\Gamma})$ and $\varepsilon_c(\textbf{k}_{\textrm{M}})$ are dominated by the zero and twice frequency spectral components, consistent with the conclusions in Ref. \cite{Dong4}.
In contrast, around $\varepsilon_c(\textbf{k}_{\textrm{K}})$, the ATAS exhibits a more complex frequency distribution.
The phenomenon may arise from the fact that, near the $\textrm{K}$ point of graphene, the energy gap between the $v$ and $c$ bands is nearly zero.
As a result, the electrons can be excited from the $v$ and $c$ bands at almost any time, leading to a broader and less structured frequency response.

Figure 2(c) illustrates the ATAS $\Delta S(\omega,t_{d})$ of gapped graphene with $\Delta_g = 0.15$ a.u.
In contrast to pristine graphene, the ATAS has six identifiable fishbone structures, as marked by the dashed and dotted rectangles.
The underlying mechanism is that in gapped graphene, electrons can be excited by the X-ray pulse from both the $g_1$ and $g_2$ bands to the $c$ band (see Sec. I of the Supplemental Material for details).
Because there exist the energy difference between $\varepsilon_{cg_1}(\textbf{k}_{\textrm{M}})$ and $\varepsilon_{cg_2}(\textbf{k}_{\textrm{M}})$, (or between $\varepsilon_{cg_1}(\textbf{k}_{\textrm{K}})$ and $\varepsilon_{cg_2}(\textbf{k}_{\textrm{K}})$, as well as between $\varepsilon_{cg_1}(\textbf{k}_{\Gamma})$ and $\varepsilon_{cg_2}(\textbf{k}_{\Gamma})$), the each fishbone structure in Fig. 2(a) split into two distinct spectral branches, resulting in six fishbone structures in Fig. 2(c).

The corresponding energy-frequency map in Fig. 2(d) reveals that the ATAS of gapped graphene is still dominated by the spectral components oscillating at zero and twice the pump laser frequency.
Interestingly, the FFSC emerges in the two spectral branches highlighted by the blue rectangles, around the spectral energies $\varepsilon_{cg_1}(\textbf{k}_{M}) + \varepsilon_{g}$ and $\varepsilon_{cg_2}(\textbf{k}_{M}) + \varepsilon_{g}$, corresponding to the $g_1 \rightarrow c$ and $g_2 \rightarrow c$ transitions for the $\textrm{M}$-point electrons.

To investigate the influence of the Berry connection on the ATAS, the spectra $\Delta S^{\boldsymbol{\mathcal{A}}}(\omega,t_{d})$ calculated using Eq. (3b) are presented in Fig. 2(e).
Compared with the full ATAS, the two fishbone structures highlighted by black rectangles in Fig. 2(c) are absent.
This absence can be attributed to the vanishing of the Berry connection at the $\Gamma$ point, which therefore does not contribute to the spectra.
In contrast, the four fishbone structures corresponding to the $\textrm{M}$- and $\textrm{K}$-point electrons persist, indicating that their presence arises from the Berry connection.

Corresponding to the spectra in Fig. 2(e), the energy-frequency map $\textrm{log}_{10}[\Delta \tilde{S}^{\boldsymbol{\mathcal{A}}}(\omega,N)]$ is shown in Fig. 2(f).
Compared with Fig. 2(d), both the zero and twice frequency spectral components are significantly suppressed.
However, for the two spectral branches related to $\textrm{M}$-point electrons, the FFSC hardly decreases.
This observation suggests that, unlike the zero and twice frequency components which are closely tied to the effective electron mass at the van Hove singularities \cite{Dong4},
the FFSC in $\Delta S^{\boldsymbol{\mathcal{A}}}(\omega,t_{d})$ is governed by the Berry connection.
In contrast, although there also exist nonzero Berry connections near $\textrm{K}$ points, the FFSCs in the spectral regions highlighted by cyan rectangles are notably weak.
The underlying mechanism responsible for this suppression will be discussed in the following sections.

\subsection{Simplified model}

To analytically investigate the effect of the Berry connection on the ATAS, we approximate the two-dimensional four-band structure using a simplified model that includes only two nonequivalent $\textrm{K}$-point electrons and three nonequivalent $\textrm{M}$-point electrons, as illustrated in the inset of Fig. 2(g).
Based on this simplified model, we calculate the numerical spectra $\Delta S^{\boldsymbol{\mathcal{A}}}(\omega,t_{d})$ and the corresponding energy-frequency map $\textrm{log}_{10}[\Delta \tilde{S}^{\boldsymbol{\mathcal{A}}}(\omega,N)]$, as shown in Figs. 2(g) and 2(h), respectively.

Compared Fig. 2(g) with Fig. 2(e) (or Fig. 2(f) with Fig. 2(h)), one can find that the simplified model qualitatively reproduces the key feature of the spectra obtained from the full two-dimensional four-band model.
Specifically, Figs. 2(e) and 2(g) reveal that, for the $\textrm{K}$-point electrons, the spectral contribution from the $g_1 \rightarrow c$ transition channel is significantly stronger than that from the $g_2 \rightarrow c$ channel.
Moreover, Figs. 2(f) and 2(h) show that for the $\textrm{M}$-point electrons, the intensity of FFSC is obviously higher than that oscillating at twice the pump frequency, whereas for the $\textrm{K}$-point electrons, the second harmonic is significantly stronger than the fundamental.
In the following sections, based on this simplified model, we develop an analytical theory for $\Delta{S}^{\boldsymbol{\mathcal{A}}}$, enabling a deeper understanding of how the Berry connection influences the ATAS.

\section{Analytical study for the effect of the Berry connection on the ATAS}

\label{s3}

\subsection{Analytical expression for spectra $\Delta{S}^{\boldsymbol{\mathcal{A}}}$}

Based on the simplified model, we derive an analytical expression for spectra $\Delta S^{\boldsymbol{\mathcal{A}}}(\omega,t_{d})$.
Because the X-ray pulse is relatively short and weak, it can be approximated to a $\delta$ function $\textbf{E}_{X}(t) =\textbf{A}_{X} \delta(t)$.
The electrons can be instantaneously excited from the core bands to the $c$ band by the X-ray pulse at the moment of $t = 0$.
According to perturbation theory and Eq. (1), the density matrix elements change from
$\rho_{g_1g_1} (\texttt{\textbf{k}}_t,t < 0^{-},t_{d}) = \rho_{g_2g_2} (\texttt{k}_t,t < 0^{-},t_{d})=1$,
$\rho_{cc} (\texttt{k}_t,t < 0^{-},t_{d}) = 0$,
and
$\rho_{cg} (\texttt{k}_t,t < 0^{-},t_{d}) = 0$ to
$\rho_{g_1g_1} (\texttt{k}_t,t = 0^{+},t_{d})=\rho_{g_2g_2} (\texttt{k}_t,t = 0^{+},t_{d}) \approx 1$,
$\rho_{cc} (\texttt{k}_t,t = 0^{+},t_{d}) \approx 0$,
and
$\rho_{g_1 c} (\textbf{\textrm{k}}_t,t = 0^{+},t_{d}) \approx i \textbf{\textit{A}}_{X} \cdot \textbf{\textrm{D}}^{\textbf{k}}_{g_1c}$,
$\rho_{g_2 c} (\textbf{\textrm{k}}_t,t = 0^{+},t_{d}) \approx i \textbf{\textit{A}}_{X} \cdot \textbf{\textrm{D}}^{\textbf{k}}_{g_2c}$.

Next, the time-dependent evolution of density matrix elements is dominated by the IR laser, and one can obtain
$\rho_{g_{1}c} (\textbf{\textrm{k}}_t,t > 0^{+}, t_{d}) = i \textbf{\textit{A}}_{X} \cdot \textbf{\textrm{D}}^{\textbf{k}}_{g_{1}c} e^{-i \int^{t}_{0} ( \varepsilon_{g_{1}c}(\textbf{\textrm{k}}_{t^{\prime}}) + \textbf{E}_{I}(t^{\prime}, t_{d}) \cdot (\textbf{\textrm{D}}^{\textbf{k}_t}_{g_{1}g_{1}}-\textbf{\textrm{D}}^{\textbf{k}_t}_{cc}))dt^{\prime}} e^{- \Gamma_{0} t} $.
According to Eq. (2), the response function is calculated by
$S(\omega, t_{d}) = \sum_{\textbf{\textrm{k}}} S_{\textbf{\textrm{k}}}(\omega, t_{d})$
and
$S_{\textbf{\textrm{k}}}(\omega, t_{d}) = 2 \operatorname{Im}[r_{g_{1} c}(\textbf{\textrm{k}}, \omega, t_{d}) \tilde{E}_{X}^{*}(\omega)] + 2 \operatorname{Im}[r_{g_{2} c}(\textbf{\textrm{k}}, \omega, t_{d}) \tilde{E}_{X}^{*}(\omega)]
\propto  2 \operatorname{Im}[r_{g_{1} c}(\textbf{\textrm{k}}, \omega, t_{d})] + 2 \operatorname{Im}[r_{g_{2} c}(\textbf{\textrm{k}}, \omega, t_{d})]$.
Because $r_{g_{1} c}(\textbf{\textrm{k}}, \omega, t_{d})$ is similar to $r_{g_{2} c}(\textbf{\textrm{k}}, \omega, t_{d})$, we derive the spectra
$\Delta S^{\boldsymbol{\mathcal{A}}}_{\textbf{\textrm{k}}}(\omega,t_{d})$ caused by  $r(\textbf{\textrm{k}}, \omega, t_{d}) \equiv r_{g c}(\textbf{\textrm{k}}, \omega, t_{d})$ with $g \in \{g_1, g_2 \}$.
When $t < 0^{-}$, the time-dependent dipole is
$r (\textbf{k}, t, t_d) = 0$,
and when $t > 0^{+}$, it is
$r (\textbf{\textrm{k}}, t, t_{d}) \approx - 2 \textit{A}_{X} \vert \textbf{\textrm{D}}_{cg}^{\textbf{k}}\vert ^{2} \sin[ \int^{t}_{0} \varepsilon_{cg}(\textbf{\textrm{k}}_{t^{\prime}}) + \textbf{E}_{I}(t^{\prime},t_{d}) \cdot \boldsymbol{\mathcal{A}}(\textbf{\textrm{k}}_{t^{\prime}}) ]d t^{\prime}] e^{- \Gamma_{0} t}$.

If the IR laser is off, the time-dependent dipole is
$r^{X}(\textbf{\textrm{k}}, t) = - 2 A_{X} \vert \textbf{\textrm{D}}_{cg}^{\textbf{k}}\vert ^{2} \sin[(\varepsilon_{cg}(\textbf{\textrm{k}})t] e^{- \Gamma_{0} t}$.
The response function is $S^{X}_{\textbf{k}}(\omega) \propto 2 A_{X} \vert \textbf{\textrm{D}}_{cg}^{\textbf{k}}\vert ^{2} \frac{\Gamma_{0}}{\Gamma_{0}^{2}+[\omega- \varepsilon_{cg}(\textbf{k})]^{2}} \equiv 2 A_{X} \vert \textbf{\textrm{D}}_{cg}^{\textbf{k}}\vert ^{2} L[\omega,\varepsilon_{cg}(\textbf{k})]$.
Here, $L(\omega,x) = \frac{\Gamma_{0}}{\Gamma_{0}^{2}+(\omega-x)^{2}}$ is the Lorentzian line shape centered at $x$. (See the supplementary materials for the detailed derivation.)

When the IR laser is turned on, the response function of the electron can be evaluated by $S_{\textbf{\textrm{k}}}(\omega,t_{d})$.
Further, By artificially setting $\boldsymbol{\mathcal{A}} = 0$, we can derive the response function $S^{\boldsymbol{\mathcal{A}}=0}_{\textbf{\textrm{k}}}(\omega, t_{d})$.
Last, we evaluate the effect of the Berry curvature on ATAS by

\vspace{-0.4cm}
\begin{align}
\Delta{S}^{\boldsymbol{\mathcal{A}}}_{\textbf{\textrm{k}}}&(\omega, t_{d})=S_{\textbf{\textrm{k}}}(\omega, t_{d}) -S^{\boldsymbol{\mathcal{A}}=0}_{\textbf{\textrm{k}}}(\omega, t_{d})\nonumber\\
&\approx 2 \textit{A}_{X} \vert \textrm{\textbf{D}}^{\textbf{k}}_{cg}\vert^{2} [J_0(b)-1] L(\omega, \mathbb{E}) \nonumber\\
& + 2 \textit{A}_{X} \vert \textrm{\textbf{D}}^{\textbf{k}}_{cg}\vert^{2} J_1(b) L(\omega, \mathbb{E} + \omega_{I}) \sin (\omega_I t_d) \nonumber\\
& - 2 \textit{A}_{X} \vert \textrm{\textbf{D}}^{\textbf{k}}_{cg}\vert^{2} J_1(b) F(\omega, \mathbb{E} + \omega_{I}) \cos (\omega_I t_d) \nonumber\\
& - 2 \textit{A}_{X} \vert \textrm{\textbf{D}}^{\textbf{k}}_{cg}\vert^{2} J_1(b) L(\omega, \mathbb{E} - \omega_{I}) \sin (\omega_I t_d) \nonumber\\
& - 2 \textit{A}_{X} \vert \textrm{\textbf{D}}^{\textbf{k}}_{cg}\vert^{2} J_1(b) F(\omega, \mathbb{E} - \omega_{I}) \cos (\omega_I t_d) \nonumber\\
& - 2 \textit{A}_{X} \vert \textrm{\textbf{D}}^{\textbf{k}}_{cg}\vert^{2} J_2(b) L(\omega, \mathbb{E} + 2 \omega_{I}) \cos (2 \omega_I t_d) \nonumber\\
& - 2 \textit{A}_{X} \vert \textrm{\textbf{D}}^{\textbf{k}}_{cg}\vert^{2} J_2(b) F(\omega, \mathbb{E} + 2 \omega_{I}) \sin (2 \omega_I t_d) \nonumber\\
& - 2 \textit{A}_{X} \vert \textrm{\textbf{D}}^{\textbf{k}}_{cg}\vert^{2} J_2(b) L(\omega, \mathbb{E} - 2 \omega_{I}) \cos (2 \omega_I t_d) \nonumber\\
& + 2 \textit{A}_{X} \vert \textrm{\textbf{D}}^{\textbf{k}}_{cg}\vert^{2} J_2(b) F(\omega, \mathbb{E} - 2 \omega_{I}) \sin (2 \omega_I t_d).
\end{align}
Here, $F(\omega,x) = \frac{\omega - x}{\Gamma_{0}^{2} + (\omega - x)^{2}}$ is the Fano line shape centered at $x$.
In Eq. (5), $b = \mathcal{A}_x(\textbf{\textrm{k}}){E_{I0}}/{\omega_{I}}$ (see TABLE SI in the Supplemental Material for the values of parameters $b$), in which $\mathcal{A}_x(\textbf{\textrm{k}})$ is the $x$ component of $\boldsymbol{\mathcal{A}} (\textbf{\textrm{k}}) = \textrm{\textbf{D}}^{\textbf{k}}_{cc} - \textrm{\textbf{D}}^{\textbf{k}}_{gg}$.
$\mathbb{E} = \varepsilon_{cg}(\textbf{\textrm{k}}) + \xi(\textbf{\textrm{k}})$
with $\xi(\textbf{\textrm{k}}) =  {A_{I0}^{2}} \nabla^{2}_{\textrm{k}_{x}} \varepsilon_{cg}(\textbf{\textrm{k}}) / {4} $.
$J_{n}(x)$ is the $n$th-order Bessel function, corresponding to the $n$th-order harmonic components of spectra.
Equation (5) applies to the five electrons in the simplified model, while the parameters $b$, $\mathbb{E}$, and the transition matrix elements $\textrm{\textbf{D}}^{\textbf{k}}_{cg}$ are different among them.

\subsection{Analytical results and discussions}

\subsubsection{Analytical results of the fishbone structure induced by the Berry connection for the $\textrm{M}$-point electrons}

\begin{figure}[t]
\begin{center}
{\includegraphics[width=8.5cm,height=7cm]{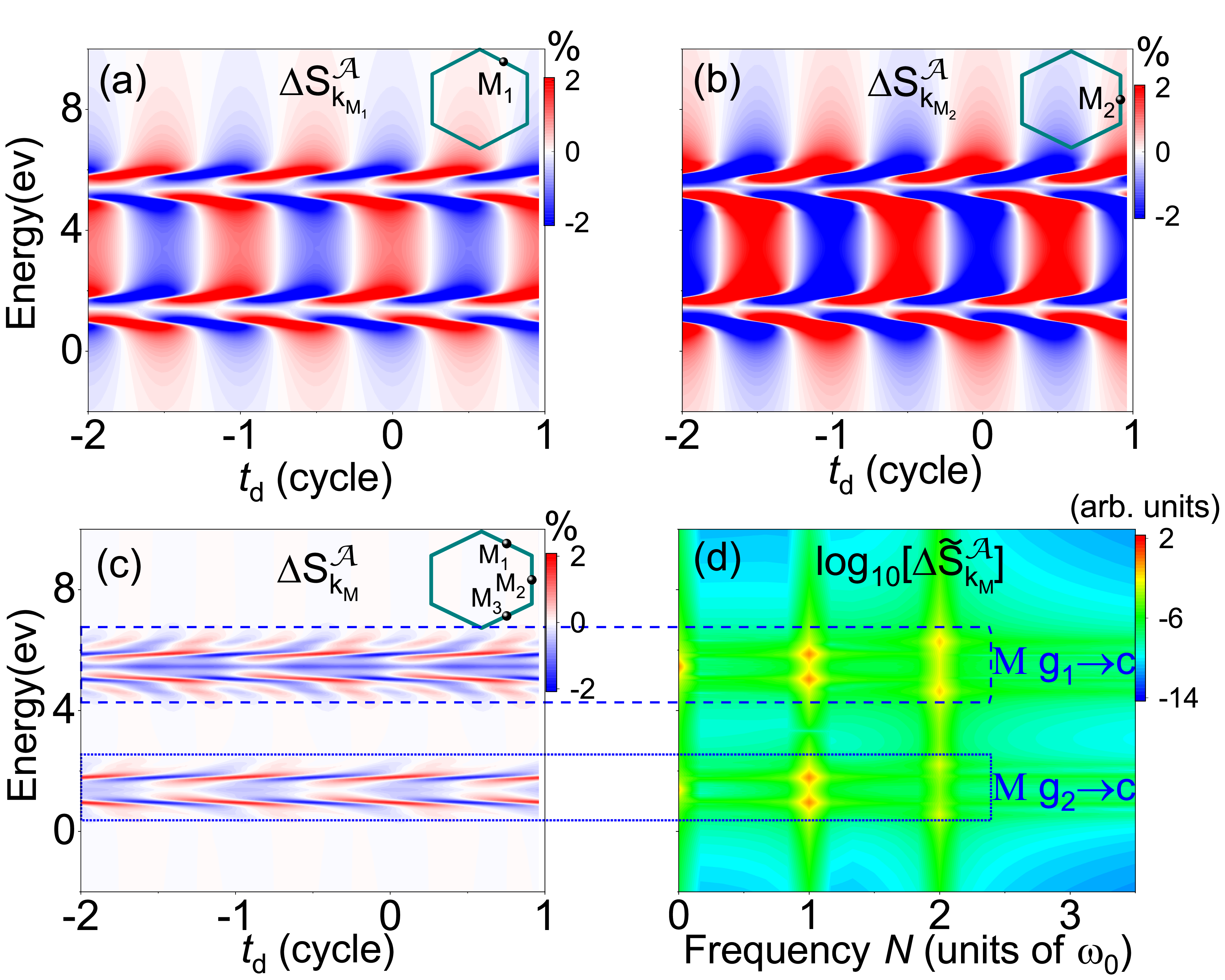}}
\caption{
(a) Analytical $\Delta{S}^{\boldsymbol{\mathcal{A}}}_{\textbf{\textrm{k}}_{\textrm{M}_1}}(\omega, t_{d})$ calculated by Eq. (5) for the electron located at $\textrm{M}_{1}$ point.
(b) Same as (a), but for the $\textrm{M}_{2}$ point.
(c) Total analytical spectra $\Delta{S}^{\boldsymbol{\mathcal{A}}}_{\textbf{\textrm{k}}_{\textrm{M}}} = \Delta{S}^{\boldsymbol{\mathcal{A}}}_{\textbf{\textrm{k}}_{\textrm{M}_1}} + \Delta{S}^{\boldsymbol{\mathcal{A}}}_{\textbf{\textrm{k}}_{\textrm{M}_2}} + \Delta{S}^{\boldsymbol{\mathcal{A}}}_{\textbf{\textrm{k}}_{\textrm{M}_3}}$.
(d) Frequency-energy map corresponding to the results in (c).
The dashed rectangle (or dotted rectangle) marks the spectrum arising from the transition channel $g_1 \rightarrow c$  (or $g_2 \rightarrow c$).
}
\label{fig:graph1}
\end{center}
\end{figure}

Figures 3(a), 3(b) and 3(c) present the analytical spectra $\Delta{S}^{\boldsymbol{\mathcal{A}}}_{\textbf{\textrm{k}}_{\textrm{M}_{1}}}$,
$\Delta{S}^{\boldsymbol{\mathcal{A}}}_{\textbf{\textrm{k}}_{\textrm{M}_{2}}}$,
and
$\Delta{S}^{\boldsymbol{\mathcal{A}}}_{\textbf{\textrm{k}}_{\textrm{M}}} = \Delta{S}^{\boldsymbol{\mathcal{A}}}_{\textbf{\textrm{k}}_{\textrm{M}_{1}}} + \Delta{S}^{\boldsymbol{\mathcal{A}}}_{\textbf{\textrm{k}}_{\textrm{M}_{2}}} + \Delta{S}^{\boldsymbol{\mathcal{A}}}_{\textbf{\textrm{k}}_{\textrm{M}_{3}}}$, respectively, calculated using Eq. (5).
These spectra consist of two fishbone structures arising from the electron transition channels $g_1 \rightarrow c$ and $g_2 \rightarrow c$.
We first focus on the upper fishbone structures in Figs. 3(a) and 3(b), which arise from the transition channel $g_1 \rightarrow c$ in $\Delta{S}^{\boldsymbol{\mathcal{A}}}_{\textbf{\textrm{k}}_{\textrm{M}_{1}}}$ (equivalent to $\Delta{S}^{\boldsymbol{\mathcal{A}}}_{\textbf{\textrm{k}}_{\textrm{M}_{3}}}$)
and
$\Delta{S}^{\boldsymbol{\mathcal{A}}}_{\textbf{\textrm{k}}_{\textrm{M}_{2}}}$.
At the point $\textrm{M}_1$ (or $\textrm{M}_3$), the parameter $b_{cg_1}^{\textrm{M}_1} =  0.0755$ (or $b_{cg_1}^{\textrm{M}_3} =  0.0755$),
while for the point $\textrm{M}_2$, $b_{cg_1}^{\textrm{M}_2} = -0.1510$.
Given that
$\vert J_1(b_{cg_1}^{\textrm{M}_1}) \vert \gg \vert J_2(b_{cg_1}^{\textrm{M}_1}) \vert$
and
$\vert J_1(b_{cg_1}^{\textrm{M}_2}) \vert \gg \vert J_2(b_{cg_1}^{\textrm{M}_2}) \vert$ in Eq. (5), the terms related to $\cos (\omega_I t_d)$ and $\sin (\omega_I t_d)$ dominate the spectra, therefore, the fishbone structures are primarily governed by the FFSC.
As a result, the oscillation frequency of the spectra in Figs. 3(a) and 3(b) match that of the driving IR laser.
Furthermore, since $\vert J_1(b_{cg_1}^{\textrm{M}_1}) \vert  <  \vert J_1(b_{cg_1}^{\textrm{M}_2}) \vert$, the intensity of the fishbone structures in Fig. 3(a) is weaker than that in Fig. 3(b).

Next, we focus on the total fishbone structure spectra
$\Delta{S}^{\boldsymbol{\mathcal{A}}}_{\textbf{\textrm{k}}_{\textrm{M}}}$
that correspond to $g_1 \rightarrow c$ transition in Fig. 3(c).
Since the energy shifts of the components $\Delta{S}^{\boldsymbol{\mathcal{A}}}_{\textbf{\textrm{k}}_{\textrm{M}_{1}}}$ and $\Delta{S}^{\boldsymbol{\mathcal{A}}}_{\textbf{\textrm{k}}_{\textrm{M}_{3}}}$ is zero, i.e., $\xi (\textbf{\textrm{k}}_{{\textrm{M}_1}}) = \xi (\textbf{\textrm{k}}_{{\textrm{M}_3}}) = 0$,
the center energy of fishbone structures is $\mathbb{E}_{cg_1}^{\textrm{M}_1}=\mathbb{E}_{cg_1}^{\textrm{M}_3} = \varepsilon_{cg_1}(\textbf{\textrm{k}}_{\textrm{M}})$.
In contrast, for the point $\textrm{M}_2$,
the center energy of the spectra $\Delta{S}^{\boldsymbol{\mathcal{A}}}_{\textbf{\textrm{k}}_{\textrm{M}_{2}}}$ is $\mathbb{E}_{cg_1}^{\textrm{M}_2}= \varepsilon_{cg_1}(\textbf{\textrm{k}}_{\textrm{M}}) + \xi(\textbf{\textrm{k}}_{{\textrm{M}_2}})$ with $\xi(\textbf{\textrm{k}}_{{\textrm{M}_2}}) = 0.0015$ a.u.
For the total spectra
$\Delta{S}^{\boldsymbol{\mathcal{A}}}_{\textbf{\textrm{k}}_{\textrm{M}}} = \Delta{S}^{\boldsymbol{\mathcal{A}}}_{\textbf{\textrm{k}}_{\textrm{M}_{1}}} + \Delta{S}^{\boldsymbol{\mathcal{A}}}_{\textbf{\textrm{k}}_{\textrm{M}_{2}}} + \Delta{S}^{\boldsymbol{\mathcal{A}}}_{\textbf{\textrm{k}}_{\textrm{M}_{3}}}$,
because the sign of $J_1(b_{cg_1}^{\textrm{M}_1})$ and $J_1(b_{cg_1}^{\textrm{M}_3})$ is opposite to that of $J_1(b_{cg_1}^{\textrm{M}_2})$ (note that $J_1(b_{cg_1}^{\textrm{M}_1}) + J_1(b_{cg_1}^{\textrm{M}_3}) \approx - J_1(b_{cg_1}^{\textrm{M}_2})$),
there is destructive interference among the FFSCs.
Due to the small energy shift between $\mathbb{E}_{cg_1}^{\textrm{M}_1}$ and $\mathbb{E}_{cg_1}^{\textrm{M}_2}$, the first-order harmonic contributions do not fully cancel out.
In contrast, the second-order harmonic components are constructively enhanced as $J_2(b_{cg_1}^{\textrm{M}_1})$, $J_2(b_{cg_1}^{\textrm{M}_2})$, and $J_2(b_{cg_1}^{\textrm{M}_3})$ all have the same sign.

\begin{figure}[t]
\begin{center}
{\includegraphics[width=8.5cm,height=7cm]{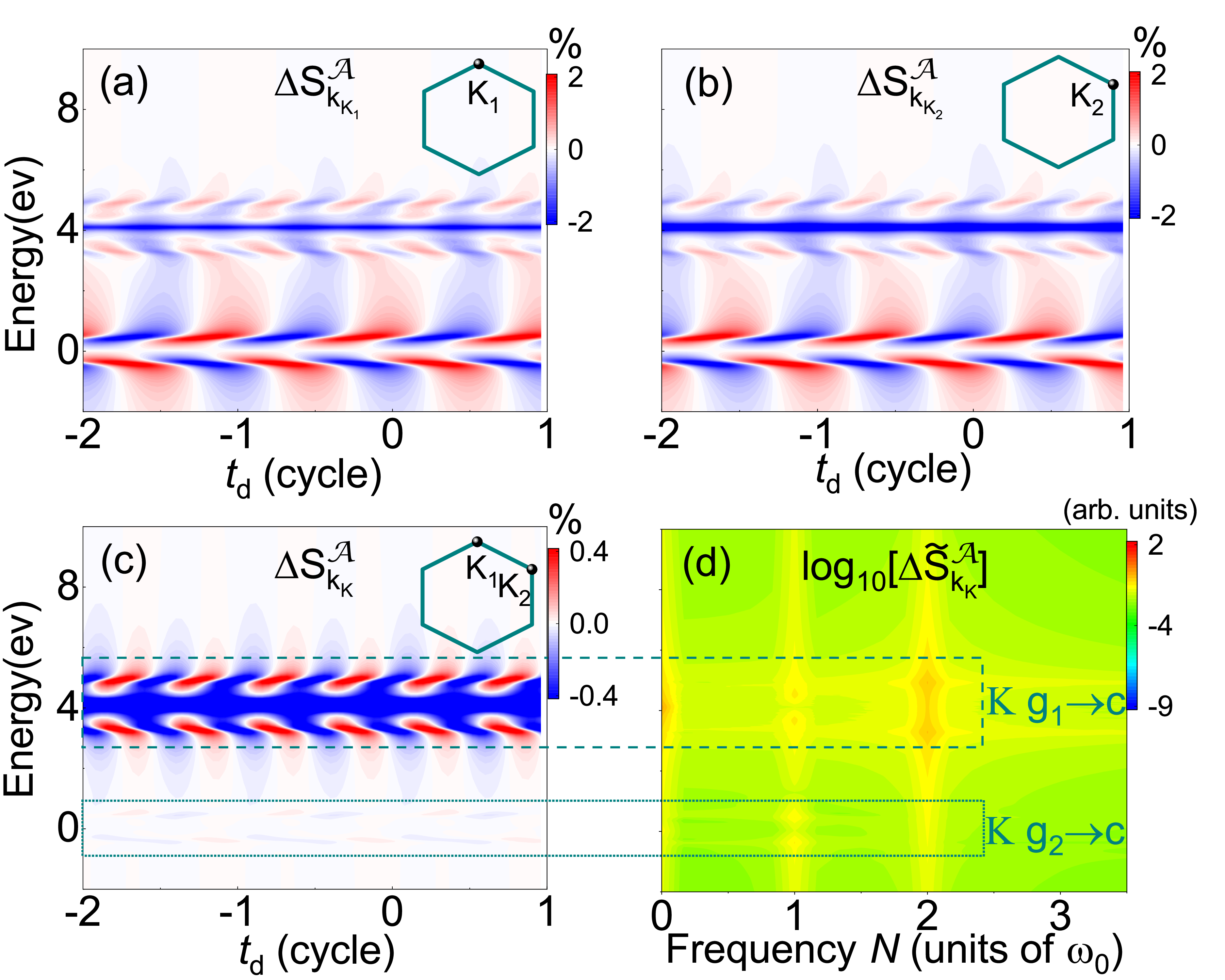}}
\caption{
(a) Analytical spectra $\Delta{S}^{\boldsymbol{\mathcal{A}}}_{\textbf{\textrm{k}}_{\textrm{K}_1}}(\omega, t_{d})$ calculated by Eq. (5).
(b) Analytical spectra $\Delta{S}^{\boldsymbol{\mathcal{A}}}_{\textbf{\textrm{k}}_{\textrm{K}_2}}(\omega, t_{d})$.
(c) Total analytical spectra $\Delta{S}^{\boldsymbol{\mathcal{A}}}_{\textbf{\textrm{k}}_{\textrm{K}}} = \Delta{S}^{\boldsymbol{\mathcal{A}}}_{\textbf{\textrm{k}}_{\textrm{K}_1}} + \Delta{S}^{\boldsymbol{\mathcal{A}}}_{\textbf{\textrm{k}}_{\textrm{K}_2}}$.
(d) Frequency-energy map corresponding to the total spectra in (c).
The dashed rectangle (or dotted rectangle) marks the spectrum arising the transition channel $g_1 \rightarrow c$ (or $g_2 \rightarrow c$).
}
\label{fig:graph1}
\end{center}
\end{figure}

For the fishbone structures arising from the transition channel $g_2 \rightarrow c$, the Bessel function arguments are $b_{cg_2}^{\textrm{M}_1} = b_{cg_2}^{\textrm{M}_3} = - 0.0189$ and $b_{cg_2}^{\textrm{M}_2} = 0.0378$.
Since the signs of $J_1(b_{cg_1}^{\textrm{M}_1})$ and $J_1(b_{cg_2}^{\textrm{M}_1})$ (as well as $J_1(b_{cg_1}^{\textrm{M}_2})$ and $J_1(b_{cg_2}^{\textrm{M}_2})$) are opposite, the fishbone structures corresponding to the $g_2 \rightarrow c$ channel appear with opposite phase relative to those from the $g_1 \rightarrow c$ channel, i.e., the upper and lower structures in Fig. 3(a) (or Fig. 3(b)) exhibit opposite behavior.
Furthermore, similar to the $g_1 \rightarrow c$ case, destructive interference occurs among the FFSCs, while the second-harmonic components undergo constructive interference due to consistent sign of $J_2(b_{cg_2}^{\textrm{M}_1})$, $J_2(b_{cg_2}^{\textrm{M}_2})$, and $J_2(b_{cg_2}^{\textrm{M}_3})$.

Corresponding to the total spectra $\Delta{S}^{\boldsymbol{\mathcal{A}}}_{\textbf{\textrm{k}}_{\textrm{M}}}$ in Fig. 3(c),
the energy-frequency map is shown in Fig. 3(d).
The results indicate that although destructive and constructive interference occur at the fundamental and second harmonics, respectively, the FFSC still dominates in intensity.
These observations are consistent with the results in Figs. 2(f) and 2(h).

\subsubsection{Analytical results of fishbone structure induced by the Berry connection for the $\textrm{K}$-point electrons}

Next, we turn our attention to the fishbone structures related to the $\textrm{K}$-point electrons.
Figures 4(a) and 4(b) display the analytical spectra
$\Delta{S}^{\boldsymbol{\mathcal{A}}}_{\textbf{\textrm{k}}_{\textrm{K}_{1}}}$ and
$\Delta{S}^{\boldsymbol{\mathcal{A}}}_{\textbf{\textrm{k}}_{\textrm{K}_{2}}}$, respectively.
Corresponding to the transition channel $g_2 \rightarrow c$, the spectral oscillation frequency agrees with that of the IR laser, as $\vert J_1(b_{cg_2}^{\textrm{K}_1}) \vert \gg \vert J_2(b_{cg_2}^{\textrm{K}_1}) \vert$
and
$\vert J_1(b_{cg_2}^{\textrm{K}_2}) \vert \gg \vert J_2(b_{cg_2}^{\textrm{K}_2}) \vert$.
However, due to the sign inversion $J_1(b_{cg_2}^{\textrm{K}_1}) \approx - J_1(b_{cg_2}^{\textrm{K}_2})$,
combined with the fact that
$\mathbb{E}_{cg_2}^{\textrm{K}_1} = \mathbb{E}_{cg_2}^{\textrm{K}_2}$,
the FFSCs interfere destructively.
As a result, the total intensity of the first-order harmonic is significantly suppressed, as seen in Figs. 4(c) and 4(d).

Corresponding to the transition channel $g_1 \rightarrow c$, because of the small values of $J_1(b_{cg_1}^{\textrm{K}_1}) $ and $J_1(b_{cg_1}^{\textrm{K}_2})$,
the spectra are dominated by the twice frequency spectral components as shown in Figs. 4(a) and 4(b).
In Fig. 4(c), the total analytical spectra
$\Delta{S}^{\boldsymbol{\mathcal{A}}}_{\textbf{\textrm{k}}_{\textrm{K}}} = \Delta{S}^{\boldsymbol{\mathcal{A}}}_{\textbf{\textrm{k}}_{\textrm{K}_{1}}} + \Delta{S}^{\boldsymbol{\mathcal{A}}}_{\textbf{\textrm{k}}_{\textrm{K}_{2}}}$ show constructive interference in the second harmonic components
owing to the approximate equality $J_2(b_{cg_1}^{\textrm{K}_1}) \approx J_2(b_{cg_1}^{\textrm{K}_2})$.
The corresponding energy-frequency map, displayed in Fig. 4(d), further confirms the dominance of the twice frequency spectral components, consistent with the numerical results shown in Figs. 2(f) and 2(h).

\subsubsection{Discussion about the dependence of the fundamental frequency components of the ATAS on the energy gap}

\begin{figure}[t]
\begin{center}
{\includegraphics[width=8.5cm,height=7cm]{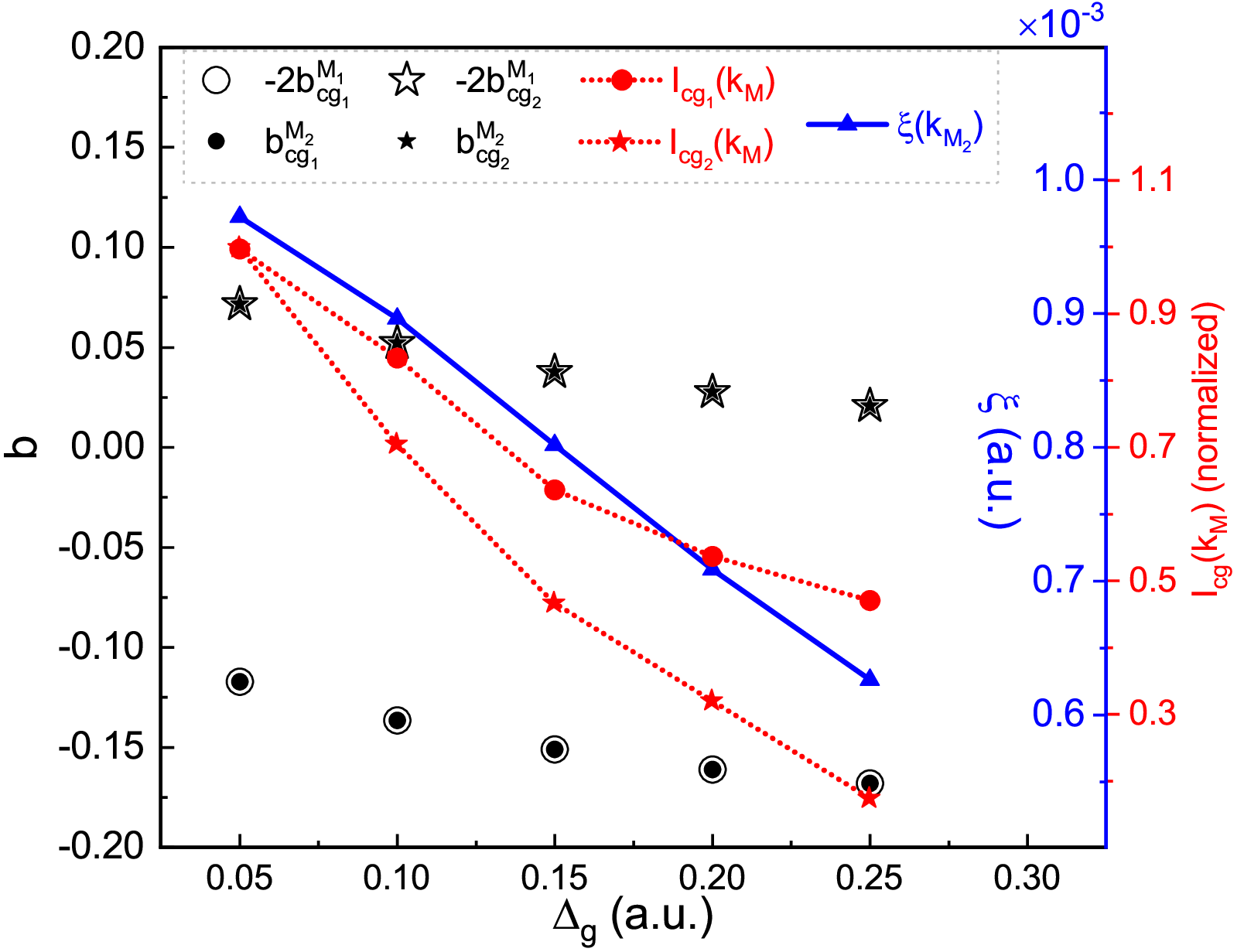}}
\caption{
Parameters $b$ as a function of $\Delta_g$ plotted by the scatters, corresponding to the transition channels $g_1 \rightarrow c$ and $g_2 \rightarrow c$ for the electrons located at $\textrm{M}_1$ and $\textrm{M}_2$ points.
The red dotted curves are the intensity of the first-order harmonic component  $I(\textbf{k}_{\textrm{M}})$, which are calculated by Eq. (6) and normalized.
The blue solid curve is the energy shift $\xi(\textbf{k}_{\textrm{M}_2})$.
}
\label{fig:graph1}
\end{center}
\end{figure}

From the above discussions, it can be concluded that for the spectra related to the $\textrm{M}$-point electrons, the intensity of FFSC depends not only on the value of Berry connections but also on the shift energy $\xi(\textbf{\textrm{k}}_{{\textrm{M}_2}})$.
In the following, we investigate how the intensity of the FFSC in the ATAS varies with the Berry connection and the shift energy, for different energy gaps in our gapped graphene model.

In Fig. 5, the scatter plots show the parameters $b$, which are directly related to the Berry connections, as a function of the energy gap $\Delta_g$.
Moreover, we evaluate the intensity of the FFSC related to the $\textrm{M}$-point electrons for the transition channels $g_1 \rightarrow c$ and $g_2 \rightarrow c$, using the equation
\begin{align}
I_{cg_1(g_2)}(\textbf{k}_{\textrm{M}}) = \int_{\varepsilon_{l}}^{\varepsilon_{u}} \Delta \tilde{S}(\omega,N=1) d \omega,
\end{align}
in which the integration limits are defined as $ \varepsilon_{l} = \varepsilon_{cg_1(g_2)}(\textbf{k}_{\textrm{M}}) + \varepsilon_{g} - 2\omega_0 $ and $ \varepsilon_{u} = \varepsilon_{cg_1(g_2)}(\textbf{k}_{\textrm{M}}) + \varepsilon_{g} + 2\omega_0 $.
Furthermore, the shift energy $\xi(\textbf{k}_{\textrm{M}_2})$ is illustrated by the blue point curve in Fig. 5.
Note that as mentioned above, $\xi(\textbf{k}_{\textrm{M}_1}) = \xi(\textbf{k}_{\textrm{M}_3}) = 0$ in our model.
As the energy gap increases, $\xi(\textbf{k}_{\textrm{M}_2})$ decreases monotonically, implying that the destructive interference becomes increasingly significant.

For the transition channel $g_2 \rightarrow c$, the absolute value of the parameter $b$ decreases with increasing $\Delta_g$, indicating that the amplitude of the FFSC in the spectra gradually diminishes.
Consequently, the total first-order harmonic yield $I_{c g_2}$ also decreases, as shown in Fig. 5.
In contrast, for the transition channel $g_1 \rightarrow c$, the absolute value of the parameter $ b $ increases with the energy gap, suggesting an enhancement in the amplitude of FFSC.
However, the total yield $I_{c g_1}$ still decreases.
This apparent contradiction can be attributed to the increasing effect of destructive interference.
As $\xi(\textbf{k}_{\textrm{M}_2})$ decreases, the phase mismatch between different contributions becomes smaller, making destructive interference more effective in suppressing the overall FFSC.

\section{Conclusion}
\label{s4}

In summary, we investigate the effect of Berry connections on the ATAS in gapped graphene by numerically solving the density matrix equations based on a developed four-band model.
In contrast to pristine graphene, our numerical results exhibit the FFSC in the spectra, which is attributed to the presence of Berry connections in gapped graphene.
To gain further insight into this interesting phenomenon, we approximate the two-dimensional four-band structure using a simplified model that includes only the nonequivalent $\textrm{K}$- and $\textrm{M}$-point electrons.
The numerically simulated spectra from both the full and the simplified model show qualitative agreement.
Using the simplified model, we derive an analytical expression for the ATAS arising from the Berry connection.
Our analytical results indicate that the intensity of FFSC not only depends on the Berry connections but also relates to the energy shifts that are associated with the effective mass of electrons at the $\textrm{K}$ and $\textrm{M}$ points.
We further investigate the dependence of the FFSC related to the $\textrm{M}$-point electrons on the energy gap.
The results indicate that as the energy gap increases, the intensity of the FFSC is significantly suppressed.

\section*{ACKNOWLEDGMENTS}

This work is supported by NSAF (Grant No. 12404394, 22303023), Hebei Province Optoelectronic Information Materials Laboratory Performance Subsidy Fund Project (No. 22567634H), the High-Performance Computing Center of Hebei University.

\end{document}